# Free-electron tomography of few-cycle optical waveforms

Yuya Morimoto[1,2*], Bo-Han Chen[1,3†], and Peter Baum[1,3*]

[1] *Ludwig-Maximilians-Universität München, Am Coulombwall 1, 85748 Garching, Germany*

[2] *Ultrashort Electron Beam Science RIKEN Hakubi research team, RIKEN Cluster for Pioneering Research (CPR), RIKEN Center for Advanced Photonics (RAP), Wako 351-0198, Japan*

[3] *University of Konstanz, Universitätsstraße 10, 78457 Konstanz, Germany*

[†]*Current address: Institute of Photonics Technologies (IPT) 101, Section 2, Kuang-Fu Road, Hsinchu 300044, Taiwan*

∗Corresponding authors:

Y. M. (email: yuya.morimoto@riken.jp) P. B. (email: peter.baum@uni-konstanz.de)

Dated: November 14, 2021

**Abstract**

**Ultrashort light pulses are ubiquitous in modern research[1–19], but the electromagnetic field of the optical cycles is usually not easy to obtain as a function of time. Field-resolved pulse characterization requires either a nonlinear-optical process[20–24] or auxiliary sampling pulses that are shorter than the waveform under investigation[25–33], and pulse metrology without at least one of these two prerequisites is often thought to be impossible. Here we report how the optical field cycles of laser pulses can be characterized with a field-linear sensitivity and no short probe events. We let a free-space electron beam cross with the waveform of interest. The randomly arriving electrons interact by means of their elementary charge with the optical waveform in a linear-optical way and reveal the optical cycles as the turning points in a time-integrated deflection histogram on a screen. The sensitivity of the method is only limited by the emittance of the electron beam and can reach the level of thermal radiation and vacuum fluctuations. Besides overturning a common belief in optical pulse metrology, the idea also provides practical perspectives for in-situ characterization and optimization of optical waveforms in higher-harmonics experiments, ultrafast transmission electron microscopes, laser-driven particle accelerators, free-electron lasers or generally any experiments with waveform-controlled pulses in a vacuum environment.**

**Main**

Ultrashort light pulses with a controlled electromagnetic waveform are central to modern research, because such pulses provide access to the time-dependent response of complex materials on the level of a single cycle of light with a time resolution of attoseconds or less[1,2]. For example, few-cycle pulses can control chemical reactions[12] and photoionization processes[13], produce high-order harmonic radiation[14–18], trigger electron emission from nanostructures[3,4,19], shape free-electron beams into attosecond pulses[5,6], steer and visualize field-induced currents in complex materials[7–9] and accelerate particles with laser light[10,11]. Also, a substantial amount of energy can be concentrated into a sub-cycle time interval, providing unprecedented field strengths and intensities for rapid material processing or high-field experiments[10,11].

Essential for almost all such experiments with few-cycle light is a profound knowledge of the electromagnetic waveform in space and time, in order to relate the measured observables to the forces that produce them. It is a common belief in optics that almost any pulse characterization method requires either a nonlinear-optical process or a probe pulse of even shorter duration than the pulse to be determined, or both. For example, pulse characterization techniques like intensity autocorrelation[20], frequency-resolved optical gating (FROG)[21,22], spectral phase interferometry for direct electric-field reconstruction (SPIDER)[23] or dispersion-scan (d-scan)[24] rely on a nonlinear-optical effect such as second-harmonic generation, difference frequency mixing, higher-order harmonic generation or optical-parametric conversion. Methods like electro-optic sampling[25,26], photoelectron streaking[27,28], petahertz nonlinear optical mixing[29], all-optical cathode-ray oscilloscopes[30,31] or direct-current metrology in optoelectronic device[32,33] are all based on test pulses or controllable events that are as short or shorter than the field cycles of interest. Linear-optical methods only provide relative quantities with respect to reference light, for example with spectral interferometry[34,35]. Consequently, it is a widespread notion in ultrafast optics that a purely linear characterization of the electromagnetic field of a light wave with long probe pulses may be impossible, or at least unrealistic.

**Free-electron tomography of optical waveforms**

Here we show how an optical waveform can be determined by neither using a nonlinear interaction nor a short sampling process. The idea is based on the use of free electrons instead of photons or bound electrons as a probe. As elementary particles in free space, beam electrons are directly and immediately accelerated in the electric field of laser light via the Lorentz force and therefore offer a linear-optical and instantaneous response to the optical waveform. Figure 1a depicts our approach. The unknown optical waveform (red) and an electron beam (blue) interact with each other in an angled geometry in which the electric field of the waveform accelerates the electrons in a sideways direction. A modulation element (black), for example a mirror, provides momentum and energy conservation[6,36–39]. The temporally long electron beam continuously covers the entire optical waveform in time, and a time-integrating detector therefore records an incoherent accumulation of deflections in form of a histogram of the field cycles.

Mathematically, we describe the physics of free-electron deflection in a similar way as in attosecond photoelectron streaking[40,41]. When we approximate the electrons as point particles with a substantial fraction of the speed of light, the sideways deflection induced by a *p*-polarized plane-wave field is expressed as (see Methods)

$$\Delta p_x(x, t_0) = eCA_{\text{in}}(t_0 + \tau_{\text{VM}}(x)) \quad (1),$$

where $A_{\text{in}}(t)$ is the temporal part of the vector potential of the incident field, the same quantity that is obtained by photoelectron streaking[27,28]. $e$ is the elementary charge, $t_0$ is the arrival time of the electron on the foil at $x = 0$, and $\tau_{\text{VM}}(x)$ is the delay due to the electron-laser velocity mismatch[39,42]. The constant $C$ is a dimension-less coupling coefficient that depends on the angles of the experiment[39,42]. Quantum-mechanically, the electrons in our beam are wave packets extended in space $x$ and time $z/v_e - t$, and the interaction with the light field is described by modulations of the phase of the de Broglie wave[38,43–46]. The quantum-coherent streaking at the mirror is

$$\Delta p_{x,\text{QM}}(x, z - v_e t, t_0) = eC' A_{\text{in}}\left(\frac{z}{v_e} - t + t_0 + \tau_{\text{VM}}(x)\right), \quad (2)$$

where $C'$ is a dimensionless constant. Both pictures show that high-energy beam electrons are deflected by the cycles of an optical waveform as a function of their arrival time in a very similar way as the bound electrons of an atom when ionized by attosecond high-energy photons under presence of an optical field[47]. However, by replacing the bound electrons with the swift free electrons in a beam, by substituting the atomic core with our macroscopic mirror membrane, we obtain an instantaneous and field-linear light-electron interaction without the need for a short probe event (i.e., attosecond ionization). The time resolution of our waveform sampling is limited only by the time it takes for a beam electron to pass though the membrane, which is less than 0.3 fs in this experiment.

**Experiment and results**

In the experiment, the few-cycle laser pulses centered at 6.9 µm to be measured are produced by optical parametric amplification[48] and focused onto a 20-nm-thick aluminium foil under *p*-polarization (see Methods). The estimated peak field strength is 0.4 V/nm. An electron beam is generated at a kinetic energy of 70 keV and a velocity of $0.5c$ by laser-triggered photoemission and subsequent electrostatic acceleration[49]. The duration of these electron pulses is ~500 fs and the entire optical waveform is covered with electrons at a nearly constant rate. The diameter of the electron beam (100 µm) is smaller than the diameter of the laser focus (~400 µm) in order to measure a local field.

Figure 1b depicts the measurement results. The direct electron beam initially hits the screen at $x = y = 0$, but in presence of the optical field it is elongated into a line shape along the *x* axis. This streaking direction is parallel to the electric field vector of the optical waveform, and the magnitude of deflection is directly proportional to the electromagnetic field strength (Extended Data Fig. 1). Interestingly, the measured pattern shows several special points where there are either maximum intensities (for example at $P_{\text{pos}}$ and $P_{\text{neg}}$) or where there are abrupt, stepwise intensity decreases (cutoffs). The dotted lines mark four such special points at −1.2 mrad, −0.6 mrad, +0.7 mrad and +1.4 mrad. Note that the measured pattern is highly asymmetric: All cutoffs and also the four accumulation points do not have corresponding points at opposite *x*; for example, $P_{\text{pos}}$ is at 1.2 mrad and $P_{\text{neg}}$ at –1.0 mrad. The right panel of Fig. 1b shows an integration along the *y* axis into which no streaking is induced. We see again the asymmetric peaks $P_{\text{pos}}$ and $P_{\text{neg}}$ as well as the four cutoffs of the deflection signal (dotted lines).

Given the linear relation of beam deflection to the optical waveform according to Eqs. (1) and (2), electrons accumulate on the detector mainly at such deflection angles where the optical waveform has a turning point, that is, an approximately constant field as a function of time. In other words, each local maximum or minimum of the waveform produces a cutoff in the streaking pattern at a deflection value given by the local peak field strength. The measured four cutoff points (dotted lines) therefore reveal that the optical waveform is a few-cycle pulse with two particular field crests in positive direction and two particular field crests in negative direction. The intensity of $P_{neg}$ is 2.1 times higher than $P_{pos}$, suggesting that $P_{neg}$ is induced by two temporally-separated negative peaks of the optical waveform. We therefore conclude, so far without any numerical analysis, on a cosine-shaped waveform that consists of one positive maximum and two similar negative extrema at adjacent times. Although the precise timing of the field cycles is not yet determined, we see that the qualitative nature of an optical waveform including its carrier-envelope phase and sequence of cycles can be deduced.

In order to confirm this assessment, we plot in Fig. 1c the results of an electro-optical sampling measurement of our pulses[17,48]. In order to account for the sampling of a $A_{in}(t)$ in our experiment [see Eqs. (1) and (2)], we adjust the carrier-envelope phase to give a cosine-shaped waveform. The result is shown in Fig. 1c as a grey line. Indeed, the optical waveform with its asymmetric shape has a complex series of field cycles at the positions and field strengths that have been deduced (dotted lines).

**Waveform reconstruction**

A simple procedure allows to reconstruct a full waveform from the experimental raw data. In most optical experiments, the spectrum $S(\omega)$ of the unknown field can be measured accurately with a spectrometer. For example, Fig. 2e shows the spectrum of our laser pulses as obtained with Fourier-transform spectrometry (see Methods). Using $S(\omega)$ and an arbitrary spectral phase $\varphi(\omega)$, the time-domain waveform can be expressed as

$$A_{in}(t) = \text{Re}\left[\int_0^{+\infty} \frac{\sqrt{S(\omega)}}{\omega} e^{i\varphi(\omega)} d\omega\right], \qquad (3)$$

where $\omega$ is the angular frequency. Commonly, the spectral phase is Taylor-expanded around a central frequency $\omega_0$ according to

$$\varphi(\omega) = \varphi_{CE} + \varphi_{GD}(\omega - \omega_0) + \frac{\varphi_{GDD}}{2}(\omega - \omega_0)^2 + \frac{\varphi_{TOD}}{6}(\omega - \omega_0)^3 + \cdots, \qquad (4)$$

The waveform is then determined by the three leading parameters $\varphi_{CE}$ (the carrier-envelope phase), $\varphi_{GDD}$ (the group delay dispersion, GDD) and $\varphi_{TOD}$ (the third-order dispersion, TOD). $\varphi_{GD}$ describes a group delay and only shifts the entire waveform in time. More coefficients are typically not significant for few-cycle pulses but can be incorporated if required. We now retrieve the optical waveform from a single deflection histogram by fitting Eqs. (1), (3) and (4) to the measurement results. We note that a retrieval of $A_{\text{in}}(t)$ is, in principle, possible without the use of the optical spectrum, but at least $\omega_0$ has to be given in order to map the reconstructed waveform crests to timing information via the cycle period; we recall that all of our data is taken without scanning a delay time. There remains only an ambiguity in sign, to be resolved as described below.

In the measured data of Fig. 1b, a cosine-like waveform has been set up in our pulse source[48] by simply maximizing the highest positive deflection ($P_{\text{pos}}$). Therefore, we assume $\varphi_{\text{CE}} = 0$ and only optimize the two remaining parameters $\varphi_{\text{GDD}}$ and $\varphi_{\text{TOD}}$ via a least square fitting algorithm to the observed deflection signal of Fig. 1b. The red curve in Fig. 1d shows the reconstructed waveform. We see that it is nearly identical to the waveform obtained by the electro-optical sampling reference (grey) except small details at weaker cycles. Also, the reconstructed waveform corresponds well to the direct guess of the peak amplitudes from the raw data alone (dotted lines).

**Tomography scans**

In order to test the robustness and universality of our procedure, and to remove the time-axis ambiguity, we now scan the carrier-envelope phase ($\varphi_{\text{CE}}$) of the laser waveform and record a series of static, not-time-resolved electron deflection histograms as a function of $\varphi_{\text{CE}}$. Figure 2a shows in the left panel the raw data in the form of a series of deflection patterns (integrated along *y*). We stress again that no time delay is scanned, and each column in Fig. 2a is a single time-integrated image for a particular $\varphi_{\text{CE}}$. We call such data a tomographic scan, because we record a series of 'time-collapsed' optical waveforms under variation of $\varphi_{\text{CE}}$ in a very analogous way to X-ray tomography, where one accumulates a series of 'depth-collapsed' projection images under variation of an illumination angle. In Fig. 2a, we see several curved lines (e.g. dark blue, orange, red, dark red) with complex near-sinusoidal shapes as a function of $\varphi_{\text{CE}}$. These changes of cutoffs and peaks correspond to gradually changing field strengths of the different optical cycles. Deflection signals at $\varphi_{\text{CE}} = \pm\pi$ are nearly identical with those at $\varphi_{\text{CE}} = 0$ when flipped upside down, as expected from the change of a cosine-shaped to a minus-cosine-shaped waveform. The extreme cutoffs at maximum positive and negative deflections (dark blue) go in parallel, indicating the simultaneous loss and gain of the highest positive and lowest negative waveform crest as a function of $\varphi_{\text{CE}}$. Figure 2c shows such a scan for the case of an intentionally chirped optical waveform (see Methods). We now see more peaks and fringes, as expected from the multiple field cycles that are now involved. Also, the dependency on $\varphi_{CE}$ becomes more complex than that without the chirp. In particular, the result becomes sensitive to time reversal.

For analysing such tomographic scans, we simply optimize the three parameters $\varphi_{CE}$, $\varphi_{GDD}$ and $\varphi_{TOD}$ with a global fit. The right panel of Fig. 2a shows the simulated tomography signal and Fig. 2b depicts the reconstructed waveform. All the features are well reproduced. The reconstruction of the chirped pulse data of Fig. 2c is depicted in Fig. 2d. Also, here, the tomography image is well reproduced, and the reconstructed waveform now reveals an asymmetric shape in time due to the influences of third-order dispersion.

**Consistency checks with dispersion control**

In another experiment, we record electron beam deflection data as a function of a varying dispersion of the optical waveform, produced by introducing four different germanium windows into the optical beam path (see Methods). Each time we record a single waveform shadow at $\varphi_{CE} = 0$ like in Fig. 1a and then evaluate the optical waveform according to Eq. (4). The black circles in Figs. 2f and 2g show the obtained group delay dispersion (GDD, $\varphi_{GDD}$) and third-order dispersion (TOD, $\varphi_{TOD}$), respectively. When using a complete tomography scan under variation of $\varphi_{CE}$ for all four dispersions, followed by four global fits, we obtain the black open squares. Both analyses provide nearly the same results. All measured dispersion values scale approximately linear with the window thickness and the slopes are consistent with the literature values[50] (dotted lines).

The general similarity of all experimental and simulation data in Figs. 2a and 2c in combination with the agreement to electro-optical sampling and recorded dispersion data establish the accuracy of free-electron tomography to capture electromagnetic few-cycle waveforms of almost arbitrary shape. Essentially, we require neither a nonlinear optical process nor a short probe event.

**Multidimensional space-time tomography with electron microscopy**

Figure 3a depicts a proof-of-concept experiment for a combination of the above reported free-electron tomography with transmission electron microscopy. The spatio-temporal optical waveform to be determined is guided onto a large mirror membrane (grey) under an incidence angle of 45° in the $xy$ plane with help of a parabolic mirror. The laser focus is set to about 3 mm behind the crossing point with the electron beam in order to produce a non-trivial waveform in space and time. A large-diameter electron beam is converted into a nearly linear shape with help of a slit aperture (orange). Consequently, our detector screen measures time-integrated free-electron streaking along the *x* axis that is spatially resolved as a function of the position along the *y* axis. Post-specimen quadrupole magnets provide special resolution along the *y* axis while maintaining angular resolution in the streaking direction along *x*.

Figure 3b shows the observed series of two-dimensional tomography images as a function of $\varphi_\text{CE}$. Again, no short probe event is applied, and no time is scanned. For each *y* (horizontal direction) a waveform is reconstructed as demonstrated above. Figure 3c shows in the two lower panels the measured tomographic signals as a function of $\varphi_\text{CE}$ for two example positions $y_1$ and $y_2$. We see at least two special lines (dark blue and turquoise) with sinusoidal dependencies on $\varphi_\text{CE}$, but their minima and maxima are shifted substantially; see the upper panel of Fig. 3c. This observation is direct evidence for a change of the carrier-envelope phase as a function of position along *y*.

Each tomography dataset produces values for the overall waveform amplitude and the spectral phases of Eqs. (3)-(4) as a function of the position *y*. Because $\varphi_\text{GD}$ can never be revealed in our non-time-resolved experiment, we determine it with a scan of the macroscopic electron-laser delay at a fixed CEP and subsequent search for the maximum deflection angles. This measurement is the only data in this report where a time delay is scanned. Figure 3d shows the combined results. The upper panel shows the absolute field amplitude of the waveform. It reveals a Gaussian function that corresponds to the optical beam profile. The middle panel shows the group delay of the optical field. We see a nearly linear increase (dotted line) and a slight curvature at the beam centre. The slope of 0.5 fs/µm reveals that the laser pulses travel on the mirror surface at a surface speed of 2 µm/fs or 6.7*c*, indicating an incidence angle of $\theta_{yz} = 9°$ in the *yz* plane; see inset of Fig. 3a. The curvature of the group delay at the beam centre is attributed to the curved wave front of a converging beam outside the focus; see inset of Fig. 3a. The dispersions $\varphi_\text{GDD}$ and $\varphi_\text{TOD}$ are not substantially changing as a function of *y* (see Methods). However, the lower panel in Fig. 3d shows the carrier-envelope phase $\varphi_\text{CE}$ as a function of *y*. We see a change of ~π over the beam size and also a substantial curvature. These observations are incompatible with the Gouy phase shifts of a perfect monochromatic Gaussian beam and demonstrate the presence and significance of the focal phases[51,52] of realistic few-cycle fields that are, for example, essential for strong-field experiments[53]. More experimental results on few-cycle waveforms before, at, and after the nominal focus position are reported in Methods.

**Sensitivity**

How weak fields can still be detected, in principle? The limit is determined by the field-angle efficiency of the modulation element and the emittance of the electron beam. Very roughly, mirror membranes can produce a deflection angle up to[39] $\alpha_\text{deflect} \approx \frac{e}{m_e} \frac{\lambda_\text{L}}{2\pi c} \frac{1}{v_e} E_0$, where $E_0$ is the electric field amplitude and $\lambda_L$ is the central wavelength of the optical field. This angle $\alpha_{deflect}$ must be smaller than the intrinsic angle spread $\alpha_{beam}$ of an electron beam that is focused down to a diameter of $\sim \lambda_L$. Using the beam emittance $\epsilon$, we obtain $\alpha_{beam} = \epsilon/\lambda_L$. Setting $\alpha_{deflect} = \alpha_{beam}$ and assuming a high-quality electron beam[54] with $\epsilon \approx 2$ pm·rad and $v_e \approx 0.5\,c$, we obtain $E \approx 10^5$ V/m. Such values are not too far away from the typical electric field strength of thermal radiation (see Methods) or vacuum fluctuations[55] and correspond to optical pulses with merely fJ of pulse energy, the same or less than in recent results with the on-chip waveform sampling[33]. In principle, the emittance of an electron beam for tomography purposes can be almost arbitrarily improved with apertures or other filters at cost of electron flux, so further advancements can be conceived.

**Conclusion and outlook**

The concept and experimental results reported in this work establish the ability to measure optical waveforms without the need for either a nonlinear-optical phenomenon or an ultrafast time resolution. The origin of this capability is the elementary charge of the electron and its large velocity in a free-space beam. High-energy beam electrons can be focused down to sub-nanometre dimensions and a transmission electron microscope without particular time resolution is therefore useful for measuring electromagnetic waveforms of almost any optical wavelength in space and time. The reported unification of few-cycle laser science with free-space electron beams and non-time-resolved electron microscopy may therefore open up novel research directions in the fields of nanophotonics, metamaterials, nonlinear optics, high-field physics, free-electron quantum optics, dielectric particle acceleration and many more.

**Funding**

We acknowledge funding from the Dr. K. H. Eberle Stiftung, the Vector Stiftung, the European Research Council (CoG 647771), the German Science Foundation via SFB1432 and the FAU Emerging Talent Initiative.

**Contribution**

YM and PB conceived the concept. YM and BC performed the experiment. YM analysed the data. YM and PB wrote the manuscript with inputs from all authors.

**Acknowledgement**

We thank Simon Stork for help with mirror preparation and Ferenc Krausz for laboratory infrastructure. Y. M. acknowledges Peter Hommelhoff for general support.

**Data availability**

All relevant data are available from the corresponding authors upon reasonable request.

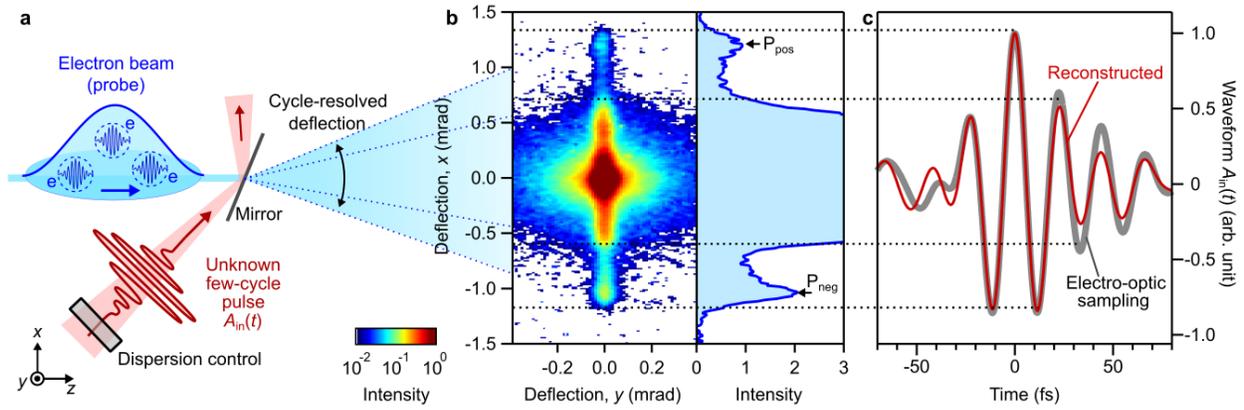

**Fig. 1. Concept and experiment for the linear-optical electron tomography of few-cycle waveforms. a**, An electron beam with a kinetic energy of 70 keV (blue) intersects with an unknown few-cycle waveform (red) at a freestanding aluminum mirror (grey). A Ge window is used for dispersion control. Depending on their random arrival time, beam electrons are accelerated sideways by the electric field cycles of the unknown waveform. **b,** A time-integrating detector resolves a deflection histogram. Left panel, raw data; right panel, one-dimensional profile. **c,** Special turning points and cutoffs (dotted lines) reveal the field cycles without the need for time resolution. The reconstructed waveform (red) agrees with electro-optic sampling data (grey).

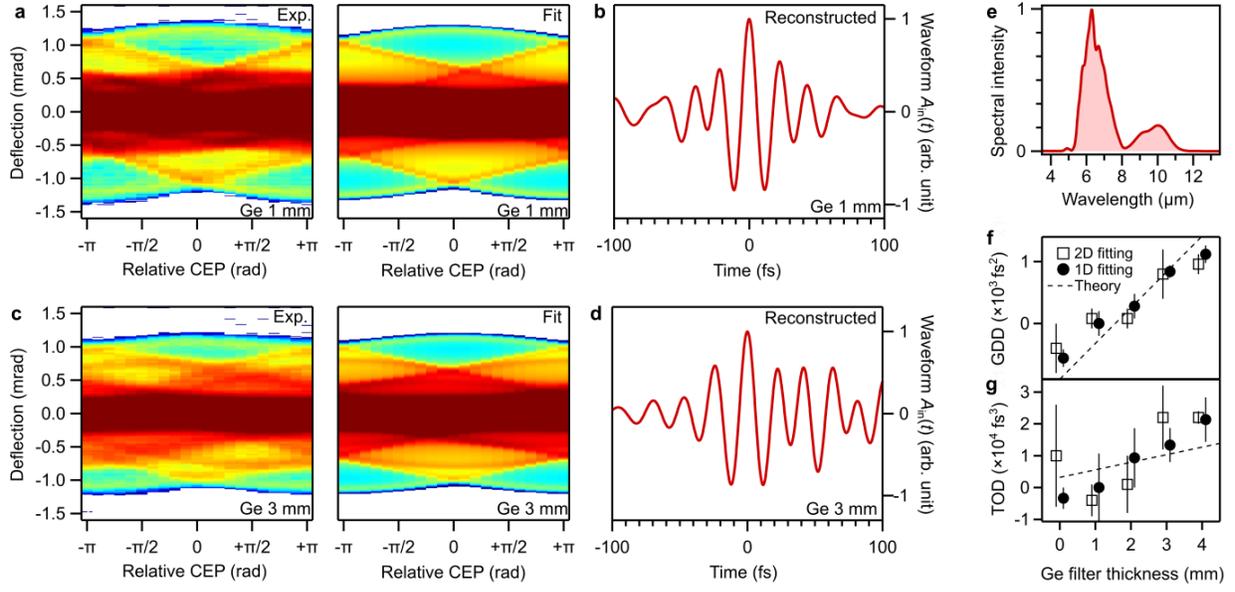

**Fig. 2. Free-electron tomography of few-cycle optical waveforms. a,** Cycle tomography of an ultrashort waveform after a 1-mm thick Ge window. Color scale, see Fig. 1. **b,** Reconstructed optical waveform. **c,** Cycle tomography results for dispersed pulses from a 3-mm-thick Ge window. **d,** Reconstructed waveform. **e,** Measured spectrum of the single-cycle mid-infrared pulse. **f,** The group delay dispersion (GDD) of reconstructed waveforms as a function of material insertion. **g,** Third order dispersion (TOD) of the reconstructed waveforms. Dotted lines in **f** and **g** are the calculated values from the refractive index of Ge. The large error bars in the TOD data indicates that the amplitudes of the most prominent field cycles in our waveform are rather independent of a varying TOD.

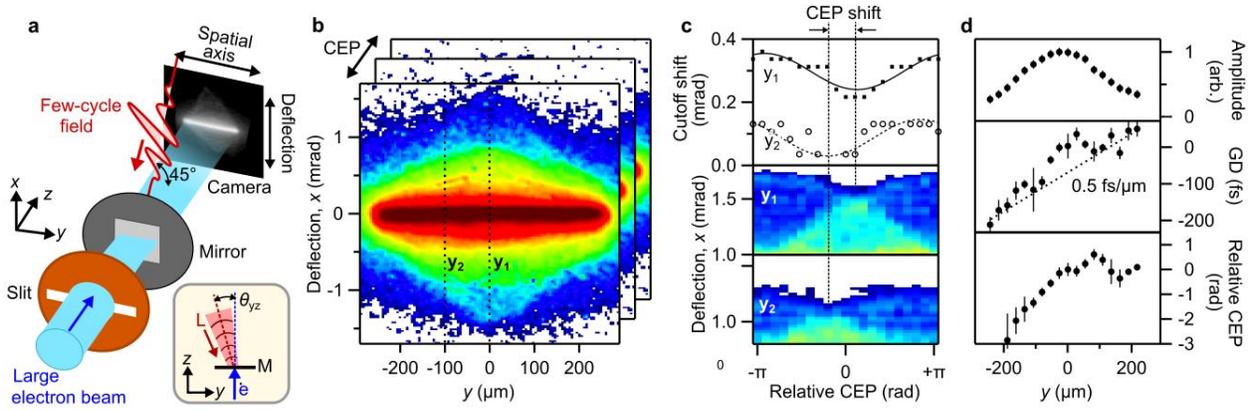

**Fig. 3. Electron microscopy combined with waveform tomography. a,** Experimental setup. A collimated electron beam (blue) passes through a slit (orange) and specimen (grey) that is illuminated by a *p*-polarized few-cycle field (red) at 45-deg incidence angle in *x-z* plane and a small angle $\theta_{yz}$ in the *y-z* plane. The optical polarization is in the $xz$ plane. Quadrupole lenses (not depicted) provide a magnified image along *y*. Inset, top view of the *y-z* plane. M, mirror; L, laser beam. **b,** Raw data as a function of the carrier-envelope phase (CEP) or delay between electron and laser beams. Color scale, see Fig. 1. **c,** Analysis of tomography lines and cutoffs. Top panel, shift of cutoff angles at $y_1 = 0$ µm (squares) and $y_2 = -100$ µm (open dots). Curves show fit results. Middle and lower panels, tomography signals around these cutoff angles. Color scale, see Fig. 1 at one tenth of the range. **d,** Evaluated waveform amplitudes, group delays and CEPs as a function of the *y* coordinate.

15